\documentclass[%
 reprint,
 amsmath,amssymb,
 aps,
 prb,
]{revtex4-2}

\usepackage{subcaption}
\usepackage{amsmath}
\usepackage{graphicx}
\usepackage{dcolumn}
\usepackage{bm}
\usepackage[justification=raggedright, format=plain]{caption}

\usepackage{natbib}
\usepackage[colorlinks=true, linkcolor=blue, citecolor=blue, urlcolor=blue]{hyperref}

\usepackage{ragged2e}  
\usepackage{etoolbox} 
\makeatletter
\long\def\@makecaption#1#2{%
  \vskip\abovecaptionskip
  \noindent
  \parbox{\linewidth}{%
    \justifying
    \hangindent=0em  
    \hangafter=1
    \setlength{\parskip}{0pt}%
    \setlength{\parindent}{0pt}%
    #1~#2%
  }%
  \vskip\belowcaptionskip
}
\makeatother

\begin{document}

\preprint{APS/123-QED}

\title{Mott Glass and Criticality in a S=1/2 Bilayer Heisenberg Model with Interlayer Bond
Dilution}

\author{Kunpeng Li}
\author{Han-Qing Wu}
\author{Dao-Xin Yao}
 \email{yaodaox@mail.sysu.edu.cn}
 \affiliation{Guangdong Provincial Key Laboratory of Magnetoelectric Physics and Devices,
 State Key Laboratory of Optoelectronic Materials and Technologies,
 Center for Neutron Science and Technology, School of Physics,
 Sun Yat-Sen University, Guangzhou, 510275, China}

\date{\today}

\begin{abstract}
We employ the stochastic series expansion quantum Monte Carlo (SSE-QMC) method to investigate the $S = 1/2$ antiferromagnetic Heisenberg model on a bilayer square lattice with diluted interlayer couplings. Both regular and random dilution patterns are considered. In systems with regular dilution, tuning the interlayer interaction drives a quantum phase transition from a N\'eel-ordered phase to a quantum disordered phase, consistent with the $O(3)$ universality class. In contrast, random dilution gives rise to a two-step transition: from the N\'eel phase to an intermediate Mott glass (MG) phase, followed by a transition to the quantum disordered phase. Within the MG phase, the uniform magnetic susceptibility exhibits a stretched-exponential temperature dependence $\chi_u \sim \exp(-b/T^\alpha)$,  $0 < \alpha < 1$. At the N\'eel-to-glass transition, quenched disorder modifies the critical exponents in a manner consistent with the Harris criterion. These findings provide new insights into disorder-driven quantum phase transitions and the emergence of glassy phases in diluted bilayer quantum magnets.

\end{abstract}

\maketitle


\section{INTRODUCTION}
\label{sec:1}
The spin-1/2 Heisenberg model on monolayer and multilayer square lattices, with various interaction patterns, has served as an important platform for studying quantum magnetism in condensed matter physics due to its rich ground-state phase~\cite{sandvik1997finite, wang2006high, huang2016quantum, PhysRevB.75.052411} and its relevance to high-temperature cuprate superconductivity~\cite{keimer1992magnetic, lee2006doping, manousakis1991spin, andersen2007disorder, barzykin1995magnetic}. Recently, bilayer quantum models have attracted renewed attention following the discovery of high-temperature superconductivity in the pressurized bilayer nickelate La$_3$Ni$_2$O$_7$~\cite{Sun2023NiSuper,luo2023bilayer,cpl_41_7_077402}. This discovery has revitalized interest in understanding exotic quantum phases within bilayer geometries, underscoring the importance of exploring generalized bilayer Heisenberg models to capture emergent many-body phenomena.
In particular, structural defects such as apical oxygen vacancies—commonly found in layered nickelate \cite{Dong2024La3Ni2O7}—may randomly suppress or remove interlayer magnetic couplings. Motivated by these considerations, we investigate bilayer Heisenberg models featuring spatially inhomogeneous or diluted interlayer interactions. Such studies may shed light on the microscopic interplay between magnetism and superconductivity in correlated bilayer systems.

Randomly suppression or removal of interlayer magnetic couplings represents a form of quenched disorder that can fundamentally alter the behavior of quantum many-body systems. In such case, the interplay between disorder and quantum fluctuations may drive the system into unconventional quantum phases~\cite{vojta2003quantum}, as demonstrated across various experimental platforms~\cite{lee1985disordered, abrahams2001metallic, fisher1989boson, lye2005bose, csathy2003substrate, yu2012quantum}. For the Heisenberg model, disorder effects have attracted sustained interest\cite{yao2010quantum,vajk2002quantum,ma2014criticality,sandvik2002classical,sandvik2006quantum, ma2015mott, sandvik2002multicritical}. In particular, bond randomness in single-layer systems has been shown to induce Mott glass (MG) phases and disorder-driven quantum phase transitions~\cite{ma2014criticality}. Similar behavior has also been reported in site- or dimer-diluted models on both single-layer~\cite{sandvik2002classical} and bilayer lattices~\cite{sandvik2006quantum, ma2015mott, sandvik2002multicritical}.

Glassy phases are commonly studied in disordered bosonic systems~\cite{iyer2012mott,yu2012bose}, where two distinct types are typically identified: the compressible Bose glass and the incompressible Mott glass~\cite{giamarchi2001competition}. These are fundamentally different from quantum disordered states in clean systems. The Mott glass phase, in particular, has long been believed to require particle–hole symmetry~\cite{prokof2004superfluid,altman2004phase}. In particular, strong-disorder renormalization group (RG) analyses and quantum Monte Carlo (QMC) simulations have shown that Mott glass phases can emerge in one- and two-dimensional Bose-Hubbard models with particle–hole symmetry~\cite{altman2004phase,iyer2012mott}. These phases are characterized by nontrivial dynamical exponents and potential experimental relevance in doped quantum magnets~\cite{yu2012bose}.
More recently, numerical studies have suggested that Mott glass–like behavior may also emerge without explicit particle–hole symmetry~\cite{wang2015anomalous}. 
In particular, spin-isotropic $S = 1/2$ systems are often mappable to bosonic models with emergent particle–hole symmetry, suggesting that disorder may induce Mott glass phases when disorder is introduced. This has been corroborated by simulations: Ref.~\cite{ma2014criticality} reported a Mott glass phase with $z = 1$ in a single-layer system with random bond modulation, and Ref.~\cite{ma2015mott} identified a Mott glass phase with $z \approx 1.36$ in a bilayer system with random dimer dilution.

Motivated by recent advances in the study of bilayer systems and Heisenberg quantum magnets\cite{ran2019criticality, liang2025magnetization, he2025magnetic,  wu2022phase,song2024quantum, wu2023classical}, we investigate the spin-$1/2$ bilayer Heisenberg model on a square lattice with dilution applied exclusively to the interlayer bonds. Unlike the dimer dilution model~\cite{sandvik2002multicritical}, which removes entire dimers—thus eliminating both spin sites along with their intra- and interlayer interactions—our approach preserves all spin sites and retains full intralayer connectivity. By selectively diluting only the interlayer couplings, we induce the formation of unconnected spin clusters confined within individual layers in some cases. This setup offers a distinct framework for exploring disorder-induced quantum phases such as the Mott glass.

\begin{figure}[t]
    \raggedright
    \begin{subfigure}[h]{0.225\textwidth} 
        \captionsetup{justification=raggedright, singlelinecheck=false} 
        \caption{}
        \raggedright
        \includegraphics[width=\textwidth]{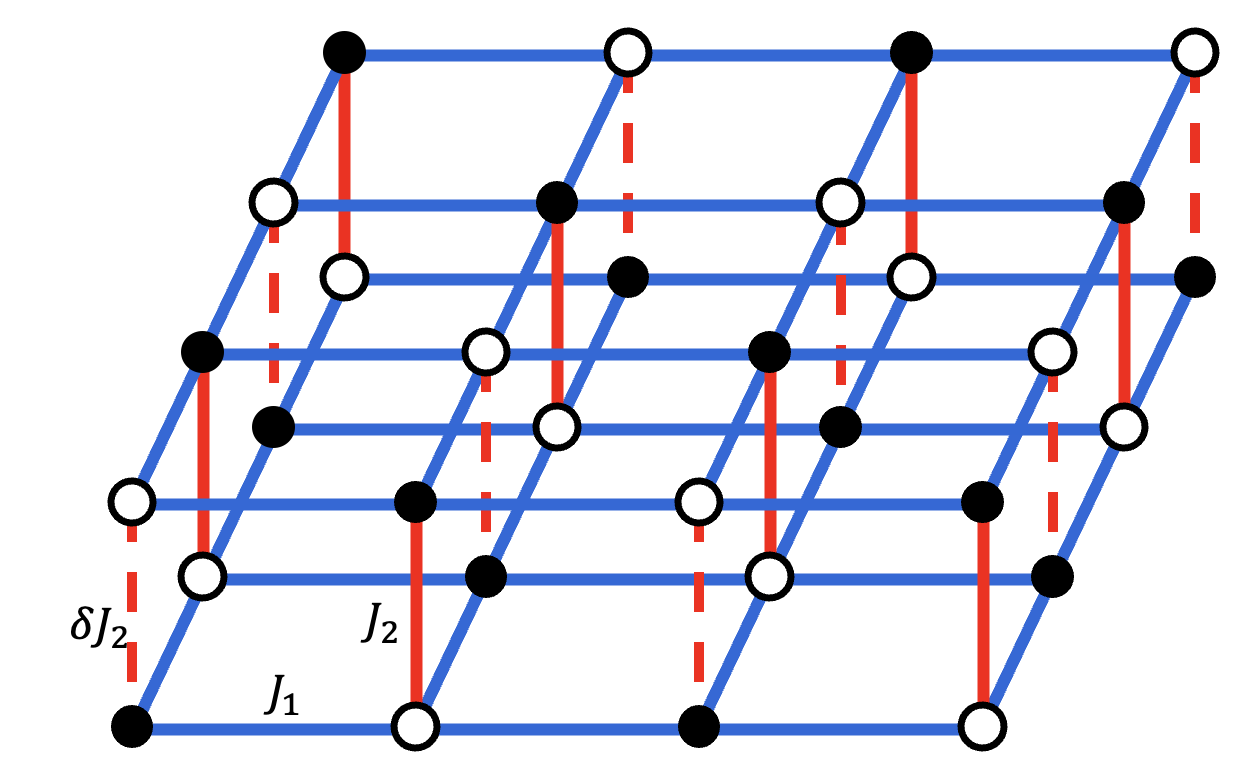}
        \label{fig:model}
    \end{subfigure}
    \begin{subfigure}[h]{0.225\textwidth} 
        \captionsetup{justification=raggedright, singlelinecheck=false} 
        \caption{}
        \raggedright
        \includegraphics[width=\textwidth]{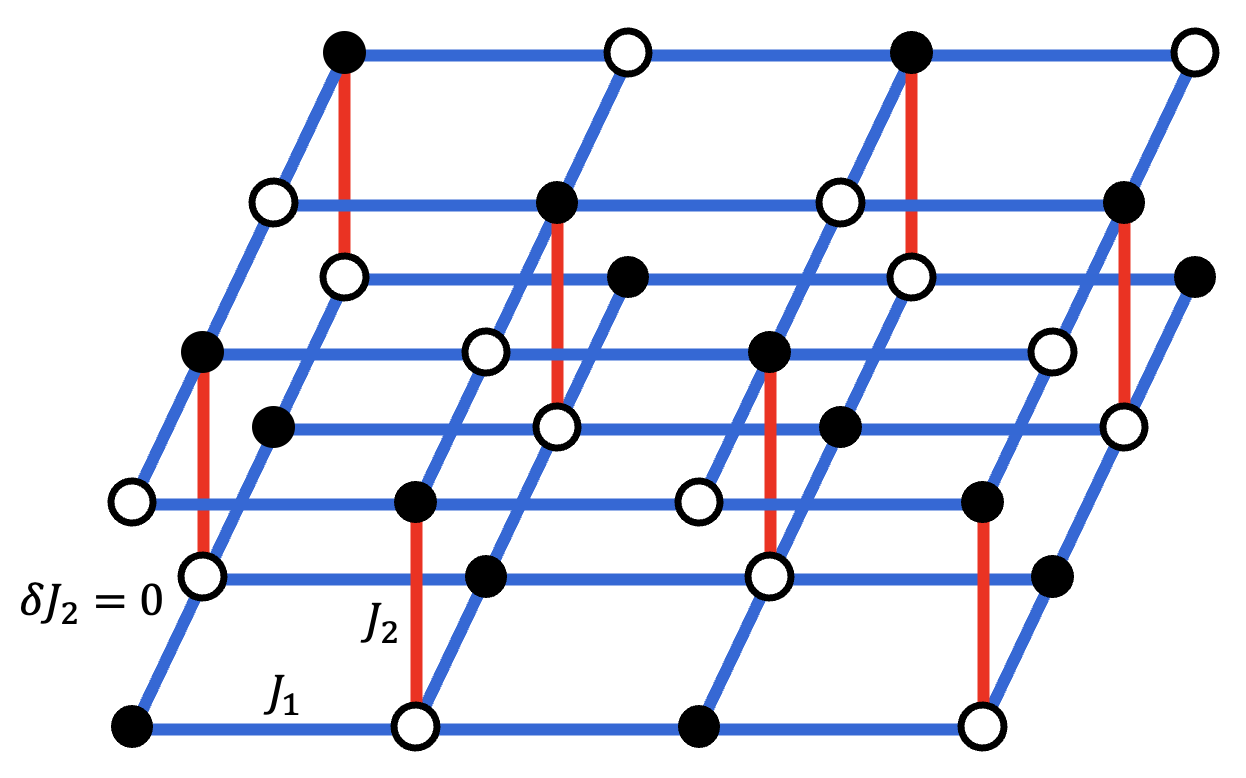}
        \label{fig:model}
    \end{subfigure}
    \vspace{0.0\textwidth} 
    
    \caption{Structure of the $S = 1/2$ bond-diluted antiferromagnetic Heisenberg bilayer with a dilution fraction of $1/2$. (a) Schematic illustration of the bilayer model, where $J_1$ denotes the intralayer coupling, and $J_2$ (solid line) and $\delta J_2$ (dashed line) denote the interlayer couplings. (b) The same model with $\delta = 0$, where the dashed interlayer bonds are fully removed.}
\end{figure}

Randomness is introduced by randomly assigned interlayer bonds with coupling $J_2$ and $\delta J_2$ with respective probabilities $1 - P$ and $P$. According to the Harris criterion~\cite{harris1974effect}, quenched disorder is irrelevant at a quantum critical point if the correlation length exponent of the clean system satisfies $\nu \geq 2/d$. Otherwise, the disorder is relevant and changes the universality class of the transition. Since disorder is introduced only in real space, the dimensionality $d = 2$ should be interpreted as the spatial dimension rather than the space-time dimension.

\begin{figure}[t]
    \begin{subfigure}[h]{0.20\textwidth}  
        \captionsetup{justification=raggedright, singlelinecheck=false} 
        \caption{}
        \centering
        \includegraphics[width=\textwidth]{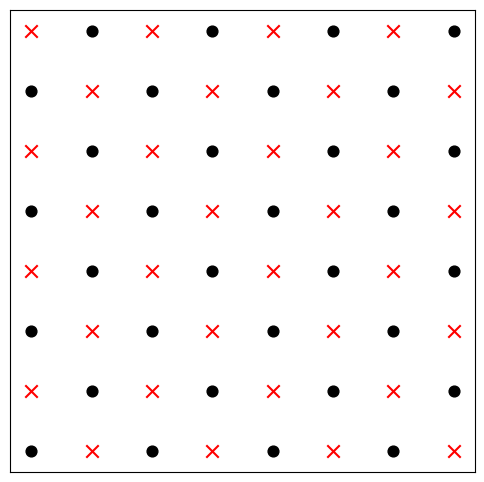} 
        \label{fig:1/2pattern}
    \end{subfigure}
    \hspace{0.01\textwidth}  
    \begin{subfigure}[h]{0.20\textwidth}  
        \captionsetup{justification=raggedright, singlelinecheck=false} 
        \caption{}
        \centering
        \includegraphics[width=\textwidth]{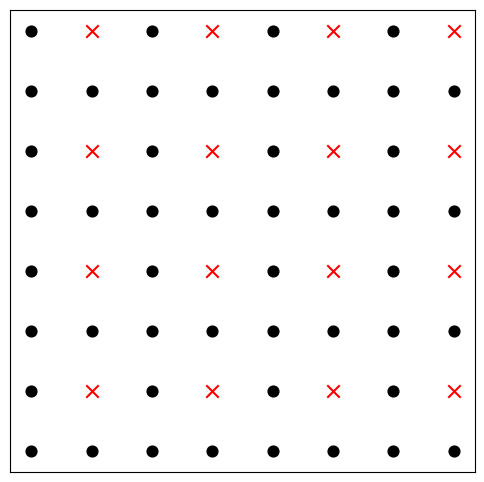}  
        \label{fig:1/4pattern}
    \end{subfigure}
    
    \begin{subfigure}[h]{0.20\textwidth}  
        \captionsetup{justification=raggedright, singlelinecheck=false} 
        \caption{}
        \centering
        \includegraphics[width=\textwidth]{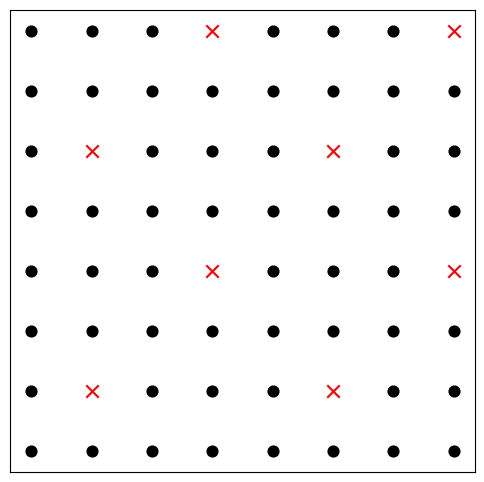} 
        \label{fig:1_8pattern}
    \end{subfigure}
    \hspace{0.01\textwidth}  
    \begin{subfigure}[h]{0.20\textwidth}  
        \captionsetup{justification=raggedright, singlelinecheck=false} 
        \caption{}
        \centering
        \includegraphics[width=\textwidth]{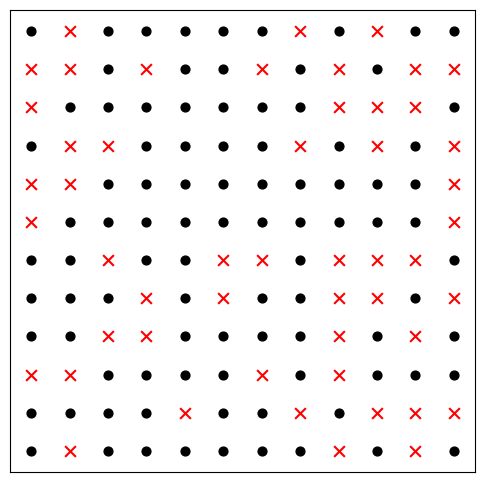} 
        \label{fig:random_pattern.}
    \end{subfigure}
    \caption{Bond-diluted bilayer lattice structures. Interlayer couplings $J_2$ and $\delta J_2$ are marked by black dot and red crosse. (a)–(c) Regular patterns with dilution fractions $1/2$, $1/4$, and $1/8$; (d) a random pattern with $P = 0.35$. We refer to each pattern by its dilution fraction hereafter.}
    \label{fig:orderpatterns}
\end{figure}

In this work, we consider both regular and random dilution patterns on the bilayer square-lattice Heisenberg model.
We first investigate the quantum phase transition in systems with regular dilution patterns to establish the universality class in the clean limit. The diluted interlayer couplings $\delta J_2$ are arranged periodically at high-symmetry positions, allowing the use of periodic boundary conditions and facilitating finite-size scaling analysis. These patterns are not only numerically tractable but may also provide experimental relevance by mimicking structural modulations.
Specifically, we consider dilution patterns with fractions of $1/8$, $1/4$, and $1/2$, as shown in Fig.~\ref{fig:orderpatterns}, to examine how both interactions and dilution density influence the critical behavior.
We then move to random dilution, where each interlayer bond is independently diluted with probability $P$, to study the effect of quenched disorder, test the Harris criterion, and explore the possible emergence of a Mott glass phase.

We employ the SSE-QMC method~\cite{sandvik2010computational} to investigate the ground-state properties of the bilayer Heisenberg model under various dilution patterns. The remainder of this paper is organized as follows: In Sec.~\ref{sec:2}, we detail the bond dilution model, introduce the physical observables considered, and outline the finite-size scaling techniques. Sec.~\ref{sec:3} presents simulation results for regular dilution configurations. In Sec.~\ref{sec:4}, we analyze randomly diluted systems and discuss the resulting phase behavior. Finally, Sec.~\ref{sec:5} summarizes our findings and provides concluding remarks.

\section{Model AND Method}
\label{sec:2}
\subsection{Halmitonian}
The Hamiltonian of the S = 1/2 antiferromagnetic bond dilution Heisenberg model can be written as
\begin{equation}
\begin{split}
H =\; & J_1 \sum_{a\in \{1,2\}} \sum_{\langle ij \rangle} S_{i,a} \cdot S_{j,a} \\
      & + J_2 \sum_{i \in \mathcal{N}} S_{i,1} S_{i,2}
        + \delta J_2 \sum_{j \in \mathcal{D}} S_{j,1} S_{j,2},
\end{split}
\end{equation}
where $\langle ij \rangle$ denotes intralayer nearest-neighbor pairs, and $a = 1, 2$ indexes the two layers. The intralayer coupling is denoted by $J_1$, and the interlayer couplings consist of regular bonds $J_2$ and diluted bonds $\delta J_2$ (with $\delta \geq 0$), located at positions belonging to the sets $\mathcal{N}$ and $\mathcal{D}$, respectively.
Throughout this work, we set the intralayer coupling $J_1 = 1$ as the energy unit and define the dimensionless coupling ratio $g = J_2 / J_1$. Our focus lies on locating the quantum critical points and determining the phase diagram of the model for various dilution configurations. In the limit $\delta = 1$, the system reduces to the conventional bilayer antiferromagnetic Heisenberg model, for which the critical coupling is known to be $g_c = 2.522$~\cite{wang2006high}.
Owing to the inherent symmetry between the normal and diluted bonds, configurations with dilution probabilities $P$ and $1 - P$ are equivalent under an exchange of $J_2$ and $\delta J_2$. This symmetry also applies to the corresponding regular dilution patterns with complementary dilution fractions. This symmetry allows the $3/4$ and $7/8$ patterns to be inferred from the $1/4$ and $1/8$ patterns via the mapping $(g_c, \delta) \rightarrow (\delta g_c, 1/\delta)$.

\subsection{Observables}
We compute three key observables—spin stiffness, Binder cumulant, and uniform magnetic susceptibility—and analyze their finite-size scaling (FSS) behavior to extract the critical points and critical exponents~\cite{luck1985corrections,jiang2012monte,wenzel2008evidence,shao2016quantum, ma2018anomalous}. These quantities provide complementary information for accurately determining the quantum critical behavior of the system.

The spin stiffness $\rho_s$ characterizes the response of the system to a twist in the boundary conditions and is defined as the second derivative of the free-energy density $f$ with respect to the twist angle $\phi$:
\begin{equation}
\rho_s = \frac{1}{N} \frac{\partial^2 f}{\partial \phi^2} = \frac{3}{4\beta} \langle w_x^2 + w_y^2 \rangle,
\end{equation}
where the winding-number estimators $w_\alpha$ ($\alpha = x, y$) are given by
\begin{equation}
w_\alpha = \frac{1}{L} \left\langle (N_\alpha^+ - N_\alpha^-) \right\rangle.
\end{equation}
Here, $N_\alpha^+$ and $N_\alpha^-$ denote the total number of off-diagonal operators $S_i^+ S_j^-$ and $S_i^- S_j^+$ in the SSE operator string~\cite{sandvik1997finite}, respectively.
At a quantum critical point, the spin stiffness follows the scaling relation
\begin{equation}
\rho_s \sim L^{2-d-z},
\end{equation}
where $d$ is the spatial dimension and $z$ is the dynamical critical exponent. Hence, the product $\rho_s L^{d+z-2}$ becomes size-independent at the critical point. 

The Binder ratio is a dimensionless quantity derived from the fluctuations of the staggered magnetization and serves as a universal indicator of quantum criticality. In this work, we use the second-order Binder ratio $R_2$, defined as
\begin{equation}
R_2 = \frac{\langle m_z^4 \rangle}{\langle m_z^2 \rangle^2},
\end{equation}
where the staggered magnetization $m_z$ is given by
\begin{equation}
m_z = \frac{1}{N} \sum_{i=1}^{N} (-1)^{x_i + y_i} S_i^z.
\end{equation}
Here, $x_i$ and $y_i$ denote the spatial coordinates of site $i$, and $N = 2L^2$ is the total number of spins in the bilayer system.
The Binder ratio is dimensionless and universal regardless of the detailed structures and couplings of the model. Because of this property, $R_2$ is widely used to determine both the location and universality class of the transition.

The uniform magnetic susceptibility $\chi_u$ provides additional information about quantum criticality and is defined as
\begin{equation}
\chi_u = \frac{\beta}{N} \left\langle \left( \sum_{i=1}^{N} S_i^z \right)^2 \right\rangle,
\end{equation}
where $\beta = 1/T$ is the inverse temperature. Close to the quantum critical point, the temperature dependence of $\chi_u$ follows the  form~\cite{fisher1989boson}:
\begin{equation}
\chi_u(T) = a + b T^{d/z - 1},
\end{equation}
where the constant $a$ vanishes precisely at the critical point. The scaling of $\chi_u(T)$ thus provides an additional means to identify the quantum critical point and extract the dynamical exponent $z$.

\subsection{Finite-size scaling}

According to the FSS theory, a physical quantity $Q$ near its critical point $g_c$ obeys
\begin{equation}
    Q(t,L) = L^{\kappa/\nu} f\bigl(tL^{1/\nu}, \lambda_1 L^{-\omega_1}, \lambda_2 L^{-\omega_2}, \dots \bigr),
\end{equation}
where $\kappa$ is the critical exponent of $Q$, $\nu$ is the correlation-length exponent, and $t=(g-g_c)/g_c$ is the reduced coupling. The set $\{\lambda_i\}$ denotes all irrelevant scaling fields with correction exponents $\{\omega_i\}$ arranged in ascending order, $\omega_{i+1}>\omega_i$. If only the leading term is retained, the quantity $L^{-\kappa/\nu} Q(t,L)$ becomes size-independent at the critical point and obey
\begin{equation}
    Q(t,L)L^{-\kappa/\nu}  = f\bigl(tL^{1/\nu} \bigr),
\end{equation}
which allows data collapse analysis to determine the quantum critical point and exponent $\nu$ accurately.

In practice, however, finite-size corrections can be significant, requiring the inclusion of subleading scaling terms in Eq.~(9) to obtain high-precision estimates of the critical point. To account for these effects, we describe the size-dependent shift of the crossing point using
\begin{equation}
g_c(L) = g_c(\infty) + a L^{-\omega},
\end{equation}
where $\omega$ is the effective correction-to-scaling exponent. This exponent can be expressed as $\omega = 1/\nu + \omega'$, where $\omega'$ captures additional subleading corrections~\cite{jiang2012monte,wenzel2008evidence,campostrini2002critical,shao2016quantum}.
We determine $g_c(L)$ by locating the crossing points of data pairs with sizes $(L/2,L)$. To estimate the statistical uncertainty of $g_c(L)$, we generate $10^3$ bootstrap samples by adding Gaussian noise with a standard deviation set by the QMC error bars.
In our analysis, we assume a common extrapolated critical coupling $g_c(\infty)$ across all observables\cite{ma2014criticality}, while allowing the correction exponents $\omega$ to vary in order to capture observable-specific subleading effects.

\section{NUMERICAL RESULTS FOR REGULAR DILUTION PATTERNS}
\label{sec:3}
\subsection{Phase boundaries of regular dilution patterns}
\begin{figure}[h]
    \centering
    \includegraphics[width=0.48\textwidth]{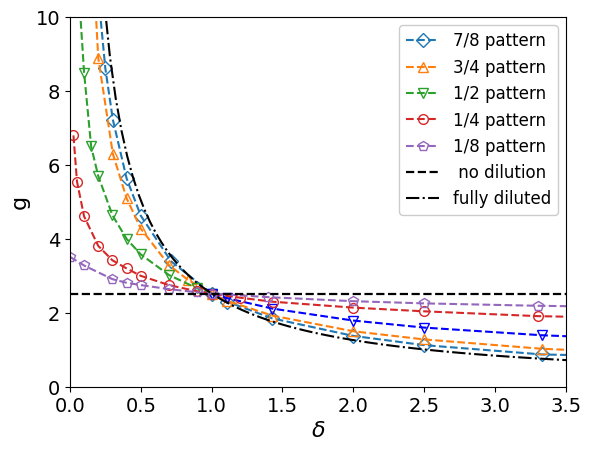}
    \caption{Phase boundaries of regular dilution patterns with dilution fractions ranging from $1/8$ to $7/8$.}
    \label{fig:diagram2}
\end{figure}

We investigate the bilayer square-lattice with periodic boundary conditions using the stochastic series expansion (SSE) quantum Monte Carlo method. To ensure statistical accuracy, we generate approximately $10^5$ samples for regular dilution patterns and $10^7$ for random case. Ground-state properties are accessed by setting the inverse temperature to $\beta = 2L$ for system sizes up to $L=64$. In this section, we focus on regular dilution configurations with diluted bonds placed on isolated sites.
We examine the quantum phase transitions and determine the corresponding phase boundaries for regular patterns with dilution fractions ranging from $1/8$ to $1/2$, as illustrated in Fig.~\ref{fig:diagram2}. In addition, by applying the mapping $(g_c, \delta) \rightarrow (\delta g_c, 1/\delta)$, we obtain the phase boundaries for the complementary $3/4$ and $7/8$ patterns, which are related to the $1/4$ and $1/8$ patterns, respectively.

In the limit $\delta = 1$, the system reduces to the clean $S = 1/2$ bilayer Heisenberg model~\cite{wang2006high}. Our simulations confirm that the phase boundaries for all regular patterns intersect at point $(\delta, g) = (1, 2.522)$. For $\delta > 1$, we observe a monotonic increase in the critical coupling $g_c$ as $\delta$ decreases, with the shift becoming more pronounced for higher dilution fractions. These results highlight that both the strength and spatial density of interlayer coupling dilution significantly influence the location of the critical point and the robustness of Néel order under partial interlayer connectivity. In addition to the isolated dilution site patterns shown in Fig.~\ref{fig:orderpatterns}, we also considered alternative configurations with the same dilution fractions. Our simulations indicate that the resulting phase boundaries exhibit only minor differences from those presented in Fig.~\ref{fig:diagram2}. Therefore, for clarity, we show only one representative phase boundary for each dilution fraction. 

 In the case of regular dilution, the phase boundary is bounded between two limiting curves. In the absence of dilution, the phase boundary is described by the clean bilayer Heisenberg model. Conversely, in the fully diluted limit where all interlayer bonds are weakened to $\delta J_2$, the model becomes equivalent to the clean case with rescaled couplings, leading to a phase boundary described by $g \delta = 2.522$.
 Since our regular dilution configurations lie between these two extremes, their critical lines fall within the region bounded by these curves. Notably, the inverse proportional curve does not intersect the vertical axis, implying that for certain highly diluted patterns, no quantum critical point exists at $\delta = 0$. This is indeed observed in patterns with dilution fraction $1/2$ or higher, where no clear transition is identified at $\delta = 0$ within the accessible coupling range. This suggests that the critical coupling $g_c$ may diverge in such cases.
The influence of dilution fraction on the location of the transition point will be further examined in the context of random dilution in Sec.~\ref{sec:4}.

\subsection{Quantum critical point and exponents}

We investigate the critical behavior of the system for all regular configurations shown in Fig.~\ref{fig:orderpatterns}, and in this section we mainly present the numerical results for the $1/8$ and $1/2$ patterns .
To accurately locate the quantum critical point and determine the critical exponents, we analyze three key observables—the spin stiffness $\rho_s$, the Binder ratio $R_2$ and the squared staggered magnetization $m_z^2$ —using finite-size scaling. 
\begin{figure}[h]
    \centering
    \includegraphics[width=0.5\textwidth]{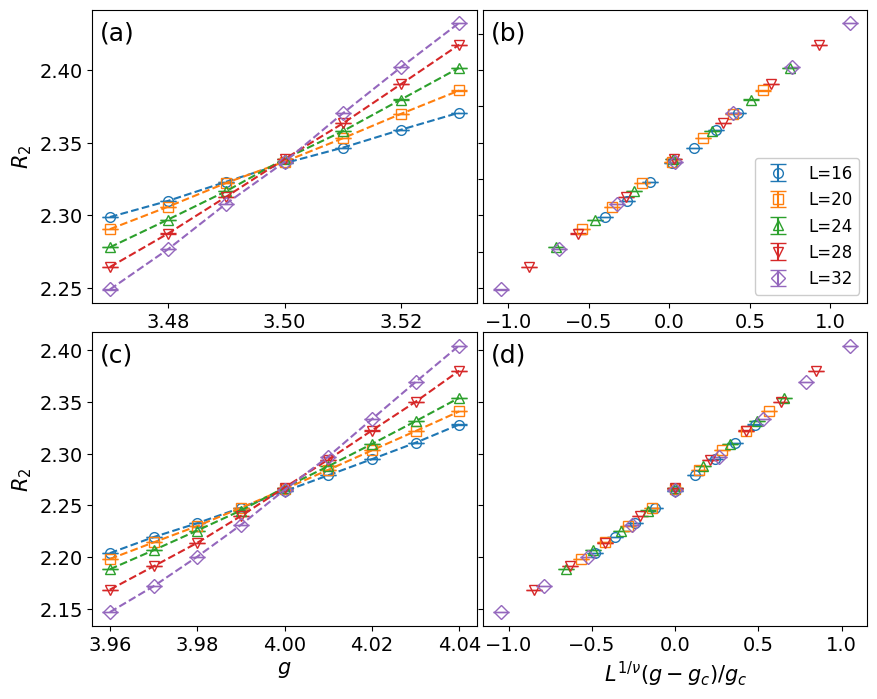} 
   \caption{Data for the regular dilution patterns with size up to L=32. (a)(b) show results for the 1/8 dilution pattern with $\delta = 0$; (c)–(d) for the 1/2 pattern with $\delta = 0.4$. (b)(d) show data collapses with fitted exponents $\nu$ = 0.716(2),0.717(3), respectively.}  
    \label{fig:result_for_8_0}
\end{figure}
For the clean bilayer Heisenberg model, one can expect $d=2$ and $z=1$~\cite{wang2006high}, leading to $\rho_s \sim L^{-1}$ becoming size-independent at the critical point. To determine the critical coupling $g_c$, we plot $\rho_s$ as a function of $g$ for various system sizes $L$ and locate the crossing points of these curves. The correlation-length exponent $\nu$ is then extracted by performing data-collapse analysis using the scaling form
\begin{equation}
\rho_s L = f(t L^{1/\nu}),
\end{equation}
where \( t = (g - g_c)/g_c \) is the reduced coupling.
The Binder ratio \( R_2 \) also exhibits a characteristic size-independent crossing near \( g_c \). Its scaling form,
\begin{equation}
R_2 = f(t L^{1/\nu}),
\end{equation}
is identical to that of \( \rho_s L \), making \( R_2 \) a reliable observable for determining both the critical point and exponent of the transition.
Additionally, we analyze the squared staggered magnetization \( m_z^2 \), which serves as the order parameter for the Néel phase.
Near the critical point, it follows the finite-size scaling form
\begin{equation}
m_z^2 L^{2\beta/\nu} = f(t L^{1/\nu}),
\end{equation}
allowing exponents \( \beta \) and \( \nu \) to be extracted simultaneously.

Representative QMC results for the $1/8$ and $1/2$ regular dilution cases with a fixed interaction ratio $\delta$ are shown in Fig.~\ref{fig:result_for_8_0}. The observables $\rho_s L$, $R_2$ and $m_z^2 L^{2\beta/\nu}$ display consistent size-dependent crossings near a common coupling $g$, indicating a reliable estimate of the critical point $g_c$.
\begin{table}[b]
\caption{\label{tab:table1}%
Critical points and exponents for the $1/2$ pattern at $\delta = 0.4$. 
}
\begin{ruledtabular}
\begin{tabular}{lccc}
\textrm{}&
\textrm{$g_c$}&
\textrm{$\nu$}&
\textrm{$\beta$}\\
\colrule
$\rho_sL$ &4.012(3) & 0.705(5) & -- \\
$R_2$ &4.001(2) & 0.717(3) & -- \\
$m_z^2 L^{2\beta/\nu}$ &4.008(5) & 0.704(3) & 0.366(4) \\
\end{tabular}
\end{ruledtabular}
\end{table}
To quantitatively determine the critical parameters, data collapses are performed by fitting these observables to third-order polynomials of the scaling variable $tL^{1/\nu}$. The optimal pair $(g_c, \nu, \beta)$ is extracted by minimizing the fitting residual. Across all studied regular dilution patterns, a consistent value of $\nu \approx 0.71$ and $\beta \approx 0.366$ are obtained, consistent with the three-dimensional $O(3)$ universality class.

However, for the $1/2$ dilution pattern at $\delta = 0.4$, noticeable drifts are observed in the crossing points of both $\rho_s L$ and $R_2$ as the system size increases. This behavior indicates the presence of substantial finite-size corrections and necessitates the inclusion of subleading scaling terms in Eq.~(11) to accurately determine the critical point.
\begin{figure}[t]
    \centering 
    \includegraphics[width=0.42\textwidth]{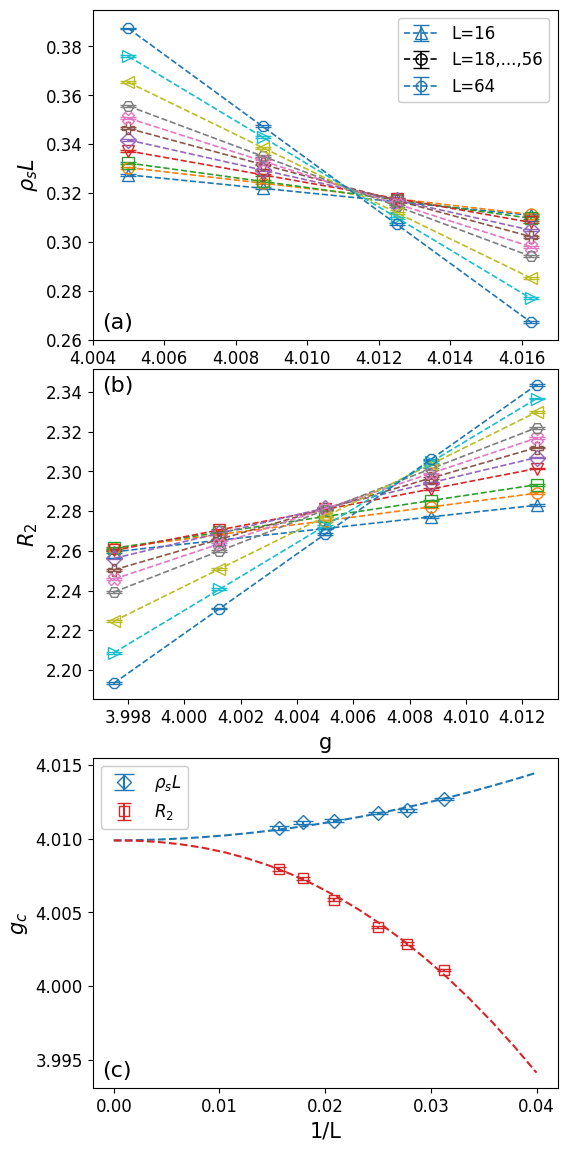}
    \caption{(a)(b)The data of spin‐stiffness and binder ratio for 1/2 dilution pattern within a narrow coupling window. (c)Finite-size estimates of the critical coupling $g_c(L)$ obtained from the $(L/2,L)$ crossing points of the spin stiffness (blue) and Binder ratio (red). The best fitting yields $g_c(\infty)$ = 4.009(1).} 
    \label{fig:data_drift} 
\end{figure}
To improve the precision of the critical coupling estimation, we restrict the fitting to narrower coupling ranges for $\rho_s L$ and $R_2$, considering system sizes up to $L = 64$. The size-dependent critical coupling $g_c(L)$ is extracted from the crossing points of $(L/2, L)$ pairs and subsequently fitted using a power-law scaling form with correction terms, as outlined previously.
To assess statistical uncertainties, we generate $10^3$ resampled datasets by adding Gaussian noise consistent with the QMC error bars. Allowing for observable-dependent correction exponents, we obtain an extrapolated critical coupling of $g_c(\infty) = 4.009(1)$, with $\omega = 1.950(1)$ for the spin stiffness and $\omega = 2.220(1)$ for the Binder ratio.

\begin{figure}[t]
    \centering 
    \includegraphics[width=0.42\textwidth]{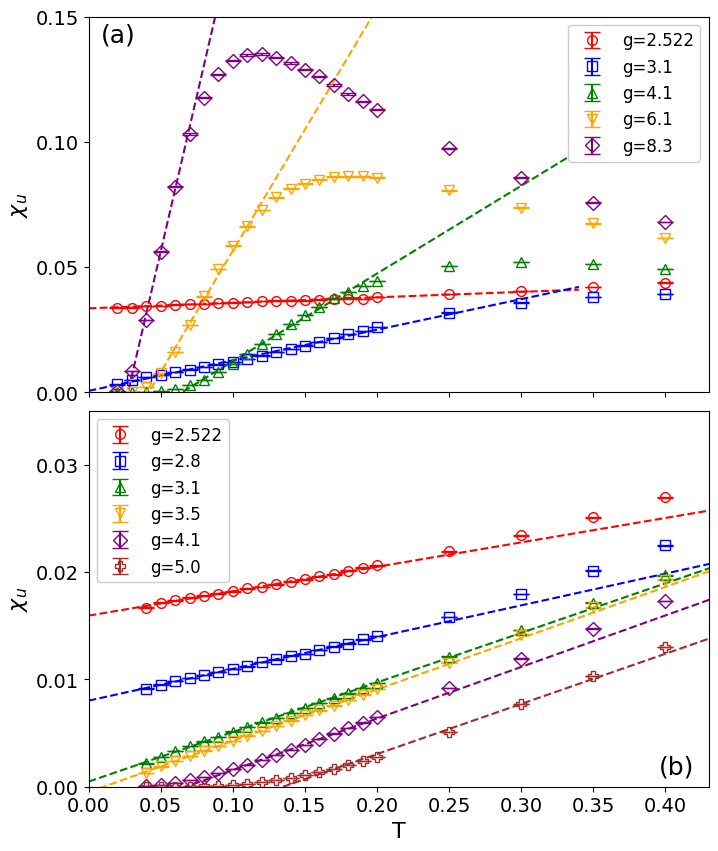}
    \caption{(a), (b) Uniform magnetic susceptibility vs the  temperature $T$ for the $1/8$ ordered dilution configuration with fixed $\delta = 0$ and $\delta = 0.4$, respectively.
} 
    \label{fig:Xu-T} 
\end{figure}

The uniform susceptibility $\chi_u$ serves as an effective probe of quantum critical behavior. To further verify the location of the critical point and the universality class of the transition, we compute $\chi_u(T)$ for the $1/8$ dilution pattern on $L = 32$ lattices, considering two interlayer coupling strengths, $\delta = 0$ and $\delta = 0.4$. The corresponding results are shown in Fig.~\ref{fig:Xu-T}.
Near the quantum critical point, $\chi_u(T)$ exhibits an approximately linear temperature dependence, consistent with a dynamical exponent $z = 1$. In particular, for $\delta = 0.4$ at $g = 3.5$, the linear fit yields a vanishing intercept as $T \to 0$, indicating that the constant term $a$ in Eq.~(8) is close to zero. This behavior suggests that $g = 3.5$ lies close to the critical coupling, consistent with earlier estimates based on the Binder ratio and spin stiffness.
Across the phase transition, $\chi_u$ displays qualitatively distinct low-temperature behavior. For $g < g_c$, it saturates to a finite value as $T \to 0$, reflecting the presence of long-range N\'eel order. Conversely, for $g > g_c$, $\chi_u$ decreases rapidly and vanishes before reaching the zero-temperature limit, signaling the emergence of a gapped disordered phase.

In contrast, when quenched disorder is present, the dynamical exponent $z$ may deviate from unity. In such case, $z$ should be treated as a free parameter to be determined through finite-size scaling and temperature-dependent analyses.

\section{NUMERICAL RESULTS FOR Random Dilution pattern with $\delta=0$}
\label{sec:4}

In the previous section, the quantum phase transition in the regularly diluted bilayer configuration was found to belong to the three-dimensional $O(3)$ universality class. Here, we turn to randomly diluted patterns characterized by varying dilution probabilities $P$.
We study randomly diluted systems with dilution probabilities ranging from 0 to 0.35, fixing $\delta = 0$ throughout. According to the Harris criterion, quenched disorder is expected to modify the critical behavior if the clean system violates the condition $\nu \geq 2/d$. In our case, the clean bilayer model exhibits $\nu \approx 0.71 < 1$ in two dimensions, suggesting that disorder would change the universality class.
\begin{figure}[h]
\centering
\includegraphics[width=0.40\textwidth]{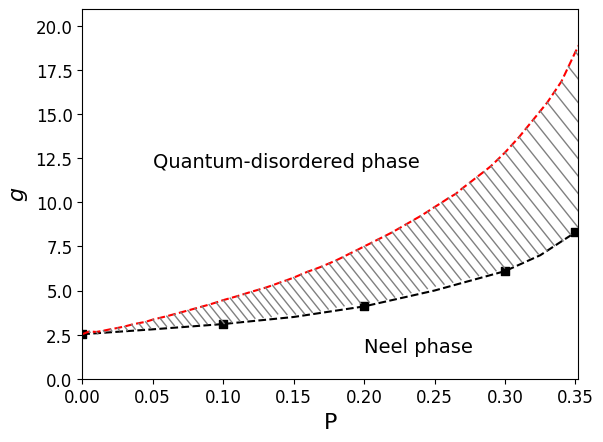}
\caption{Phase diagram of the randomly diluted configuration at $\delta = 0$, mapped in the $(P, g)$ parameter space. The shaded region denotes the Mott glass phase separating the N\'eel-ordered phase from the quantum disordered phase.}
\label{fig:phase_diagram_randomness}
\end{figure}
Evidence for this behavior has been reported in Ref.~\cite{ma2014criticality}, where the introduction of random dimer dilution lead to a change in the critical exponent $z$. Unlike their model, in which both spins and couplings are removed, our construction introduces randomness exclusively in the interlayer bonds, preserving all spin degrees of freedom. To assess the effect of such disorder, we carry out a systematic numerical investigation of the resulting quantum phase transitions.

By analyzing the Binder ratio, spin stiffness, and uniform susceptibility, we determine the phase boundaries that separate the N\'eel-ordered, Mott glass, and quantum disordered phases. The transition between the N\'eel and Mott glass phases is clearly captured by the FSS, as the two phases are distinguished by the presence or absence of long-range magnetic order. In contrast, the transition between the Mott glass and the quantum disordered phase lacks a sharp finite-size signature. Since both phases are disordered, the Binder ratio exhibits no size-independent crossing, suggesting the possible absence of a conventional second-order transition. This lack of clear FSS behavior is consistent with the underlying physical picture~\cite{giamarchi2001competition,ma2014criticality}: the Mott glass is gapless and incompressible due to rare-region effects, while the quantum disordered phase is gapped and incompressible. The transition between them likely proceeds via a broad crossover, beyond the scope of standard FSS techniques.

Moreover, we observe that the two phase boundaries rise steeply as the dilution probability approaches $P = 0.35$, making it increasingly difficult to determine the critical coupling $g_c$ with high precision beyond this point. We attribute this behavior to the proximity of the site percolation threshold, $P_c = 0.407$, above which the diluted sites are expected to form a system-spanning percolating cluster. In such clusters, interlayer couplings are effectively eliminated, and interlayer interactions cease to play a dominant role in driving the phase transition. Although determining $g_c$ becomes more challenging at high dilution levels, our results suggest that $g_c$ continues to grow with increasing $P$, and may ultimately diverge in the limit of strong dilution.

\subsection{Quantum Criticality at the N\'eel--Mott Glass Transition}

In this section, we focus on the quantum phase transition between the N\'eel and Mott glass phases, using the case with dilution probability $P = 0.2$ as a representative example. In clean systems with regular dilution patterns, the correlation-length exponent remains $\nu \approx 0.71$, indicating that the introduction of quenched disorder may modify the universality class of the transition. Consequently, the exponent $z$ and $\beta$ should not be fixed priories, but rather treated as a fitting parameter to be determined via data collapse. We begin by analyzing the Binder ratio \( R_2 \), spin stiffness \( \rho_s \) and squared staggered magnetization \( m_z^2 \) to extract the optimal set of critical parameters \( (g_c, \nu, z, \beta) \). 
\begin{figure}[h]
    \centering 
    \includegraphics[width=0.48\textwidth]{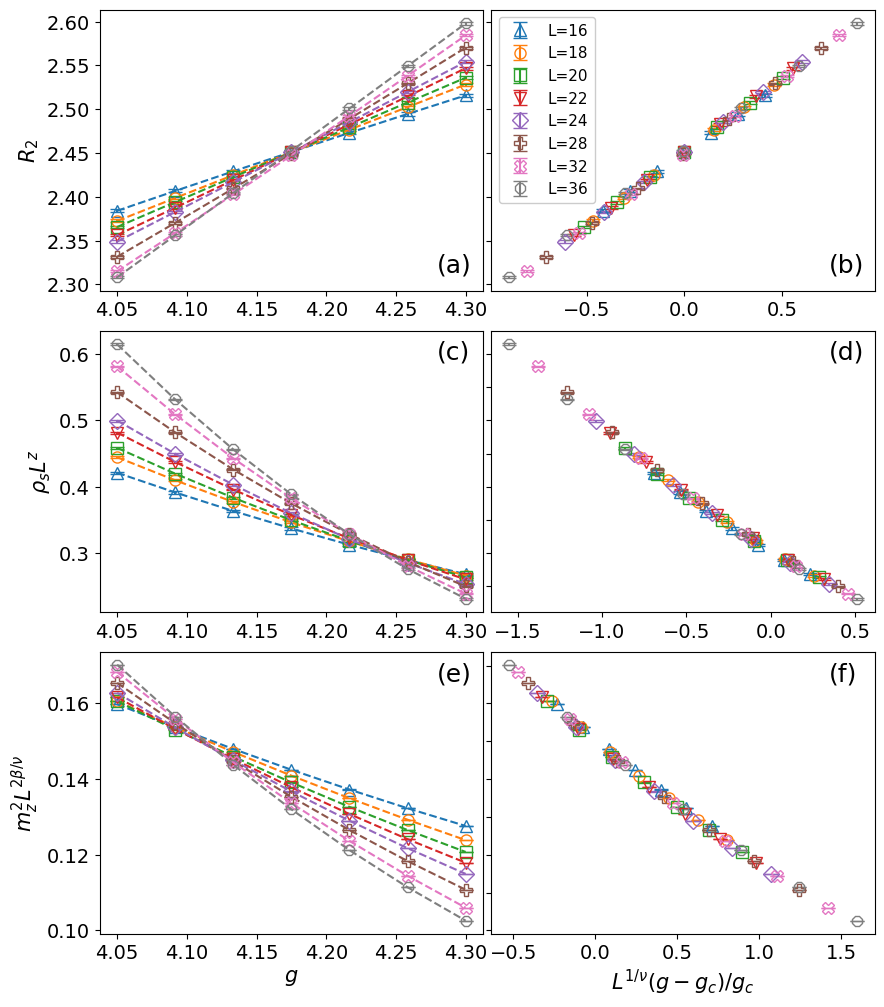}
    \caption{Scaling collapse for random dilution pattern with \( P = 0.2 \) and \( \delta = 0 \). The fitted correlation length exponent is \( \nu = 1.01(2) \), \( 1.01(0) \), and \( 1.01(0) \) for \( R_2 \), \( \rho_s \), and \( m_z^2 \), respectively. The order parameter exponent is \( \beta = 0.504(2) \), and the dynamic exponent is \( z = 1.236(1) \).} 
    \label{fig:fss_random} 
\end{figure}
\begin{figure}[t]
    \centering 
    \includegraphics[width=0.42\textwidth]{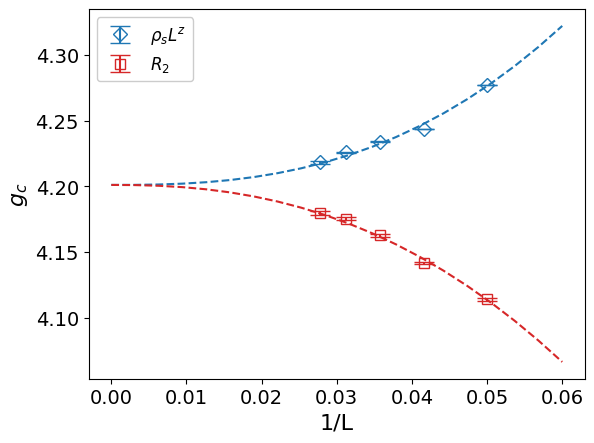}
    \caption{ Finite-size estimates of the critical coupling $g_c(L)$, obtained from the $(L/2, L)$ crossing points of the spin stiffness (blue), the Binder ratio (red) for system with $P=0.2$ and $\delta=0.4$.}
    \label{fig:drift_randomness} 
\end{figure}\textbf{}
\begin{table}[b]
\caption{\label{tab:table1}%
Critical points and exponents for the random pattern with $P=0.2$, $\delta=0.4$.
}
\begin{ruledtabular}
\begin{tabular}{lcccc}
\textrm{}&
\textrm{$g_c$}&
\textrm{$\nu$}&
\textrm{$z$}&
\textrm{$\beta$}\\
\colrule
$\rho_sL^z$ &4.23(6) & 1.01(0)& 1.236(1) &  --\\
$R_2$ &4.17(6) & 1.01(2)& -- & -- \\
$m_z^2 L^{2\beta/\nu}$ &4.11(2) & 1.01(0) & -- & 0.504(2) \\
\end{tabular}
\end{ruledtabular}
\end{table}
These parameters are determined through systematic data-collapse procedures by scanning a grid of candidate values and fitting \( R_2 \), \( \rho_s L^z \), and \( m_z^2 L^{2\beta/\nu} \) as linear functions of the scaling variable \( t L^{1/\nu} \), where \( t = (g - g_c)/g_c \) is the reduced coupling. The resulting values are further validated by examining the low-temperature scaling behavior of the uniform magnetic susceptibility.
\begin{figure}[t]
    \centering 
    \includegraphics[width=0.42\textwidth]{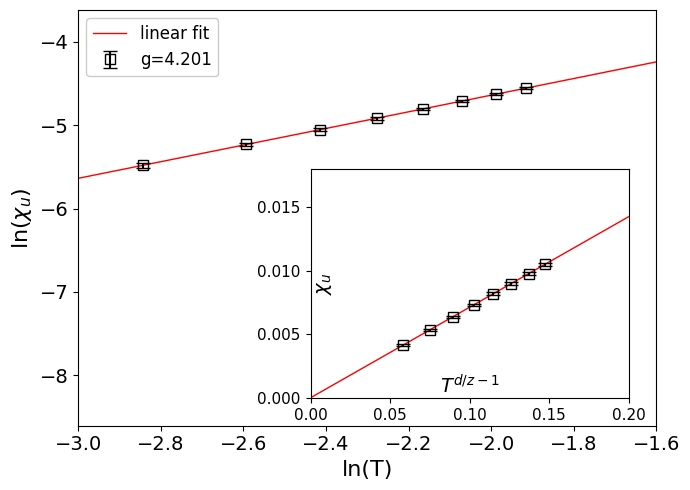}
    \caption{Temperature dependence of the uniform susceptibility $\chi_u$ at the estimated critical coupling $g = 4.201$ for L=32  with $P=0.2$.} 
    \label{fig:Xu-T_randomness} 
\end{figure}

Notably, the extracted critical exponents exhibit systematic deviations from those of the clean bilayer Heisenberg model. These discrepancies suggest that the presence of quenched disorder modifies the underlying critical behavior, potentially driving the system toward a different universality class characterized by disorder-induced scaling corrections.
To further refine the estimate of the critical coupling, we examine the finite-size drift of the crossing points obtained from $R_2$ and $\rho_s$, using the scaling form $g_c(L) = g_c(\infty) + a L^{-\omega}$. This finite-size analysis yields an extrapolated value of $g_c(\infty) = 4.201(1)$, providing a more precise and reliable reference point for subsequent thermodynamic analysis. At this critical coupling, we investigate the temperature dependence of the uniform susceptibility $\chi_u$. Near criticality, $\chi_u$ is expected to scale as $T^{2/z - 1}$, offering an independent and complementary check on the extracted dynamical exponent $z$.

Fig.~\ref{fig:Xu-T_randomness} presents the uniform susceptibility $\chi_u(T)$ at the critical coupling $g_c = 4.201$ for $L = 32$, plotted on both logarithmic and linear temperature scales. In the log-log plot, a clear linear relation is observed, from which we extract the scaling exponent $z = 1.236(6)$—in agreement with the previous estimate of $z = 1.236(1)$ obtained from spin stiffness scaling. This consistency between independent approaches reinforces the reliability and internal coherence of our critical parameter estimates.
On the linear scale, the extrapolated intercept of $\chi_u(T)$ approaches zero as $T \to 0$, indicating that $g_c = 4.201$ accurately identifies the quantum critical point for this disordered configuration without appreciable offset.

We further examine the case of $P = 0.3$, where qualitatively similar critical features persist. The extracted exponents, $\nu = 1.01(2)$, $z = 1.202(3)$ and $\beta = 0.525(5)$, exhibit a comparable level of enhancement relative to the clean bilayer case. These results lend additional support to the conclusion that the introduction of randomness systematically alters the critical exponents, suggesting a disorder-driven modification of the universality class governing the transition.

\subsection{Uniform Susceptibility and Mott Glass--Quantum Disordered Crossover}

In the preceding section, we employed finite-size scaling analysis to locate the phase boundary between the N\'eel-ordered phase and Mott glass (MG) phase. We now turn to the boundary between the MG and the quantum-disordered phase by analyzing the temperature dependence of the uniform magnetic susceptibility $\chi_u$.
In the quantum-disordered phase, $\chi_u$ is expected to exhibit activated behavior characterized by exponential suppression:
\begin{equation}
\chi_u \sim \exp\left(-\Delta / T\right),
\end{equation}
where $\Delta$ denotes the spin gap.
By contrast, previous studies~\cite{yu2005quantum,wang2015anomalous,ma2014criticality,ma2015mott} have shown that in the Mott glass phase, $\chi_u$ follows a stretched-exponential form:
\begin{equation}
\chi_u \sim \exp\left(-b / T^{\alpha}\right),
\end{equation}
with $0 < \alpha < 1$, indicating the existence of low-energy excitations associated with rare-region effects characteristic of the Griffiths regime~\cite{ma2014criticality}.
To probe this transition, we perform a detailed analysis of the temperature dependence of $\chi_u$ across a range of coupling ratios $g$, using a square lattice of linear size $L = 32$. For $g < g_c$, the uniform susceptibility $\chi_u$ approaches a finite constant as $T \to 0$, indicating a Néel-ordered phase. 
In contrast, when $g \gg g_c$, $\chi_u$ exhibits a standard exponential decay with temperature, which is typical of of the quantum-disordered phase.
However, in the regime where $g$ is slightly above $g_c$, the susceptibility follows a stretched-exponential form as described by Eq.~(16), suggesting that the system resides in a Mott glass phase. This observation indicates that the previously identified critical point marks the phase boundary between the Néel-ordered phase and quantum glass regime.

The exponent $\alpha$ in the stretched-exponential behavior increases with $g$ and approaches unity in the quantum-disordered regime. This trend enables a systematic analysis of $\alpha(g)$ to identify the phase boundary between the Mott glass and quantum-disordered phase. We find that $\alpha$ exhibits a linear dependence on $g$, which allows us to perform a linear fit and extract the critical coupling by locating the point where $\alpha = 1$. To estimate the uncertainty in the exponent $\alpha$, Gaussian noise consistent with the QMC error bars is added to the original data, and the fitting procedure is repeatedly performed on 1000 resampled datasets. The standard deviation of the resulting fitted $\alpha$ values is then used as the statistical uncertainty.

\begin{figure}[t]
    \centering 
    \includegraphics[width=0.45\textwidth]{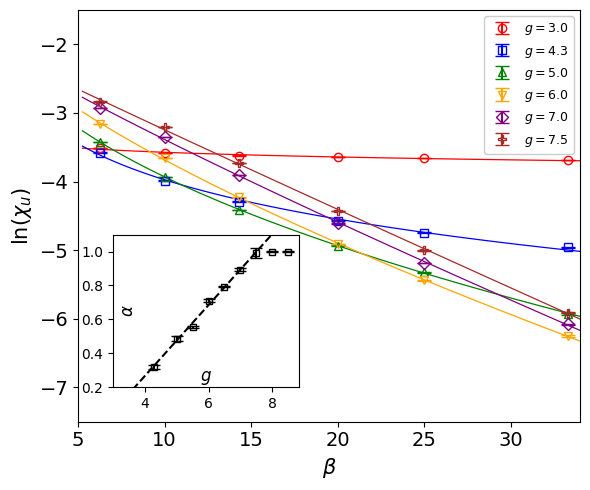}
    \caption{Inverse temperature dependence of $\log(\chi_u)$ at $\delta = 0.0$ for system with $L = 32$ and $P = 0.2$, showing stretched exponential behavior for various couplings $g$. Inset shows the extracted exponent $\alpha(g)$ with a linear fit used to determine the transition point where $\alpha = 1$.} 
    \label{fig:ln(Xu)-beta-MG} 
\end{figure}

Such stretched-exponential behavior can be naturally interpreted within a physical framework in which isolated, gapless magnetic clusters are embedded in an otherwise gapped background~\cite{ma2014criticality,wang2015anomalous}. According to Ref.~\cite{wang2015anomalous}, when the probability of a site belonging to a magnetic cluster of size $s$ decays exponentially—as is typical in sub-threshold percolation—and the excitation gap of such clusters scales algebraically with their size, the resulting uniform susceptibility $\chi_u$ exhibits a stretched-exponential temperature dependence.

In our model, the random removal of interlayer couplings with probability $P$ induces the formation of partially decoupled antiferromagnetic domains within the bilayer structure. This process gives rise to a distribution of disconnected spin clusters, especially prevalent below the percolation threshold. In this regime, the cluster-size distribution follows a power-law form, $P(s) \sim s^{-\tau}$\cite{stauffer2018introduction}, which is a hallmark of sub-threshold percolation. As emphasized in Ref.\cite{wang2015anomalous}, such a power-law distribution of cluster sizes is a crucial ingredient for the emergence of a Mott glass phase, providing a natural explanation for the observed stretched-exponential susceptibility.

\section{SUMMARY AND DISCUSS}
\label{sec:5}
We have introduced a two-dimensional antiferromagnetic Heisenberg model on a bilayer square lattice with diluted interlayer bonds. The ground-state properties and quantum phase transitions have been systematically investigated for both regular and random dilution patterns using SSE-QMC simulations, conducted on lattices up to $L = 64$ with inverse temperature $\beta = 2L$.
For the regularly diluted configurations, which represent clean systems, dilution fractions ranging from $1/8$ to $7/8$ have been considered, and the corresponding phase diagram has been constructed. The quantum phase transition in these regular patterns has been found to remain in the standard $O(3)$ universality class, characterized by critical exponents $\nu \approx 0.71$, $\beta \approx 0.366$ and $z = 1$, regardless of whether $\delta = 0$ or $\delta > 0$. In addition, the critical coupling $g_c$ has been observed to shift systematically with various dilution strength $\delta$ and dilution fraction.

We have further explored disordered systems by introducing random dilution of interlayer bonds. For instance, at $P = 0.2$ and $\delta = 0$, we have identified two distinct quantum phase transitions as the interlayer coupling ratio $g$ increases. The first transition, from the N'eel-ordered phase to a Mott glass (MG) phase, has been characterized through finite-size scaling (FSS) analyses of the Binder ratio and spin stiffness. The corresponding critical point has been determined as $g_c = 4.201(1)$, with critical exponents $\nu = 1.01(1)$, $z = 1.236(6)$ and $\beta = 0.504(2)$. These values exhibit a clear deviation from the clean-system universality class, highlighting the relevance of disorder and supporting the Harris criterion. A similar transition has also been observed at $P = 0.3$, where $\nu = 1.01(2)$, $z = 1.202(3)$ and $\beta=0.525(5)$, indicating robust trends across different disorder strengths.

Beyond the first critical point, we have found that the system enters a Mott glass phase, which is gapless yet magnetically disordered. In this intermediate regime, the uniform susceptibility $\chi_u$ has been shown to follow a stretched-exponential temperature dependence, $\chi_u \sim \exp(-b / T^\alpha)$, $0 < \alpha < 1$, reflecting rare-region effects and the absence of a true spin gap. As the interlayer coupling ratio $g$ increases further, a second transition has been identified into a gapped quantum disordered (dimerized) phase. Since both the Mott glass and quantum disordered phases lack long-range order, conventional finite-size scaling (FSS) methods are ineffective in detecting a sharp transition. To address this, we have determined the second transition point by analyzing the linear trend of $\alpha(g)$ and locating the coupling $g$ at which $\alpha$ approaches unity. This crossover behavior marks the opening of a full spin gap, where the parameter $b$ corresponds to the singlet–triplet gap $\Delta$.

Within the Griffiths-phase framework, the stretched-exponential behavior of $\chi_u$ can be naturally attributed to finite-size Néel-ordered clusters embedded within a quantum paramagnetic background. The energy gaps associated with these rare clusters decay algebraically with their size, consistent with mechanisms proposed in disordered bosonic systems~\cite{wang2015anomalous}. In our model, this behavior arises from the random removal of interlayer couplings, which leads to the formation of partially decoupled antiferromagnetic domains within the bilayer lattice.
Although studies on Mott glass behavior in $S=1/2$ systems are still limited, related works on disordered Heisenberg models~\cite{ma2014criticality, ma2015mott} reported a similar sequence of transitions—from a Néel-ordered phase to a Mott glass, and eventually to a quantum disordered state—thus supporting the generality of our findings.

In summary, by introducing bond dilution into the bilayer Heisenberg model, we have systematically investigated the quantum phase transitions under both regular and random dilution configurations using SSE-QMC simulations. Our results reveal critical behavior that is consistent with shifts in the universality class and suggest the possible emergence of a Mott glass phase. These findings may also be relevant to understanding magnetic inhomogeneity and interlayer coupling effects in nickelate superconductors, where structural defects such as apical oxygen vacancies may play a similar role to bond dilution in modifying the spin dynamics. We hope that these numerical findings will provide valuable guidance for future experiments and can be experimentally verified.

\begin{acknowledgments}
We wish to acknowledge the support of Xuyang Liang in offering suggestions. This work was supported by NKRDPC-2022YFA1402802, NSFC-92165204, NSFC-12494591, NSFC-12474248, Guangdong Provincial Key Laboratory of Magnetoelectric Physics and Devices (2022B1212010008), Guangdong Fundamental Research Center for Magnetoelectric Physics (2024B0303390001), and Guangdong Provincial Quantum Science Strategic Initiative (GDZX2401010).
\end{acknowledgments}

\appendix
\nocite{*}

\bibliographystyle{apsrev4-2}
\bibliography{apssamp}

\end{document}